# The Impact of Inter-grain Phases on the Ionic Conductivity of LAGP Solid Electrolyte Prepared by Spark Plasma Sintering

Sorina Cretu[1,2,3], David G. Bradley[4], Omer Ulas Kudu[2,3], Li Patrick Wen Feng[1], Linh Lan Nguyen[1], Tuan Tu Nguyen[2,3], Arash Jamali[2,3], Jean-Noel Chotard[2,3,5], Vincent Seznec[2,3], John V. Hanna[1,4*], Arnaud Demortière[2,3,5*], Martial Duchamp[1*]

[1] School of Materials Science and Engineering, Nanyang Technological University, 50 Nanyang Avenue, 639798, Singapore

[2] Laboratoire de Réactivité et de Chimie des solides (LRCS), Université de Picardie Jules Verne, CNRS UMR 7314, 33 rue Saint Leu, 80039 Amiens Cedex, France

[3] Réseau sur le stockage Electrochimique de l'Energie, CNRS FR 3459, 33 rue Saint Leu, 80039 Amiens Cedex, France

[4] Department of Physics, University of Warwick, CV4 7AL, UK.

[5] ALISTORE-European Research Institute, CNRS FR 3104, Hub de l'Energie, Rue Baudelocque, 80039, Amiens Cedex, France

Corresponding authors: arnaud.demortiere@cnrs.fr , martial.duchamp@gmail.com, J.V.Hanna@warwick.ac.uk

## Abstract

$Li_{1.5}Al_{0.5}Ge_{1.5}(PO_4)_3$ (LAGP) is a promising oxide solid electrolyte for all-solid-state batteries due to its excellent air stability, wide electrochemical stability window and cost-effective precursor materials. However, further improvement in their ionic conductivity performance is hindered by the presence of inter-grain phases leading to a major obstacle to the advanced design of oxide based solid-state electrolytes. This study establishes and quantifies the influence of inter-grain phases, their 3D morphology, and formed compositions on the overall ion conductivity properties of LAGP pellets fabricated under different Spark plasma sintering conditions. Based on complementary techniques, such as PEIS, XRD, 3D FIB-SEM tomography and solid-state MAS NMR coupled with DFT modelling, a deep insight into the inter-grain phase microstructures is obtained revealing that the inter-grain region is comprised of $Li_4P_2O_7$ and a disordered $Li_9Al_3(P_2O_7)_3(PO_4)_2$ phase. We demonstrate that optimal ionic conductivity for the LAGP system is achieved for the 680 °C SPS preparation when the disordered $Li_9Al_3(P_2O_7)_3(PO_4)_2$ phase dominates the inter-grain region composition with reduced contributions from the highly ordered $Li_4P_2O_7$ phases.

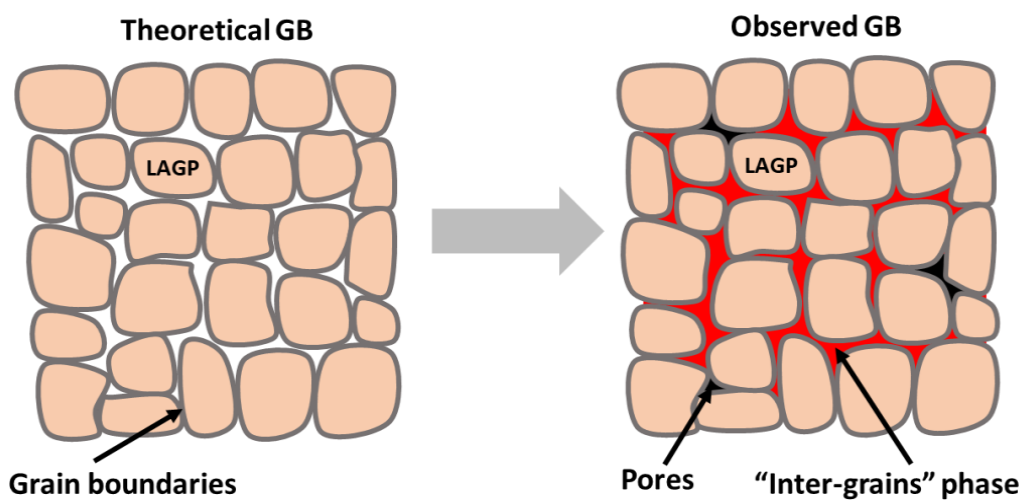

## Introduction

The rapid growth of renewable energy stimulates the development for energy storage devices with high energy density and high power. Conventional lithium-ion batteries (LiB) consisting of organic liquid electrolytes are efficient energy storage devices which are successfully commercialized since decades[1]. Despite continuous improvements, the conventional LiB technology still suffers from safety issues due to the flammability of the liquid electrolytes[2,3]. Solid state batteries (SSBs)[4–7] that use solid electrolytes (SEs)[8–17] promise to overcome this limitation by eliminating the use of flammable solvent[18–21]. SSB technology presents other advantages such as high energy density due to its compatibility with high voltage cathode materials[22,23] and opens the possibility to safely use lithium metal as the anode, thus increasing the gravimetric energy density by 40%[24,25] and simplifies the manufacturing process[26]. Within the range of available material for the SE, solid oxide electrolytes[27–32] have a high stability in air[29], wide electrochemical window[33,34] and improved chemical stability with oxide cathodes[35]; however, their implementation is limited due to their poor ionic conductivity at room temperature[36].

The superior ionic conductivity of the sulfide SEs compare to their oxide counterparts appears to be related to their reduced grain boundary resistance[37]. Different strategies have been proposed to decrease grain boundary resistances in oxide SEs, such as the introduction of $B_2O_3$ as a microstructure modifier to facilitate the grain growth and to reduce the number of grain boundaries[38], or the incorporation of a secondary lithium compounds such as $Li_3BO_3$, $Li_3PO_4$, $Li_2O$, $Li_4P_2O_7$ which segregate at the grain boundaries and subsequently inducing lower grain boundary resistance[38–40]. Nevertheless, the total ionic conductivity of oxide-based SEs, and especially $Li_{1+x}Al_{1-x}Ge_{1+x}(PO_4)_3$ (LAGP) which exhibits one of the highest ionic conductivities ($10^{-4}$ S.cm$^{-1}$ at RT) from the oxide group[32], is still an order of magnitude lower in comparison to its sulfide SEs counterparts[11,41,42]. Previous studies have demonstrated for NASICON structure-type electrolytes such as LAGP electrolytes sintered at high temperature, a partial decomposition occurs at the grain boundaries, leads to the formation of secondary phases creating an inter-grain region[43–45]. LAGP synthetized using excess Li showed an increase in total ionic conductivity due to lithium segregation at the grain boundaries which formed Li-rich structures, subsequently increasing the concentration of the charge carriers which results in an increased grain boundary conductivity[46,47]. Despite their importance, the structure and the composition of the inter-grain region are still not fully understood which stands as an obstacle for the further development of LAGP-based SSB.

To unambiguously identify the origin of the ionic conductivity losses in the inter-grain regions, this work aims to correlate the inter-grain region composition with LAGP ionic conductivity by varying the sintering temperature of LAGP in spark plasm sintered (SPS) sintered pellets. A more complete understanding of the inter-grain microstructure is obtained using complementary macroscopic techniques (compactness and impedance measurements), longer range/periodic structural techniques such as XRD, and short-range/local characterization achieved by SEM-EDX, 3D FIB-SEM tomographic reconstruction and solid-state MAS NMR coupled with materials modelling. This combined approach attempts to investigate whether specific grain boundary phase compositions can induce an improved ionic conductivity performance in the LAGP system.

## Results and Discussion

LAGP can be synthesized by various methods such as solid-state synthesis, melt quenching, microwave assisted sintering, co-precipitation, and sol-gel techniques, with the last option allowing to obtain the highest conductive solid electrolyte at the lowest sintering temperature[32,46,48–51]. In this study, LAGP powder was synthesized using sol-gel method (see the Experimental Methods section). **Figure 1a** displays the scanning electron microscopy (SEM) image of the synthesized solid electrolyte that exhibits a tetragonal morphology characterized by a particle size distribution spanning a 0.175 - 3.0 µm range with LAGP particles agglomerated in group of tens of microns (**Figure S1a**). The Rietveld refinement pattern of sol-gel synthesized solid electrolyte displayed in the **Figure 1b** shows that the main phase corresponds to LAGP (space group $R$3-c). However, secondary phases such as $GeO_2$, $Li_9Al_3(P_2O_7)_2(PO_4)_3$ (OPLA), $Li_4P_2O_7$ and $AlPO_4$ are also present, which are consistent with previous reports on the synthesis of the LAGP phase[32,52].

Oxides solid electrolytes require high temperatures to get densified and to possess highly ionic conductivity[27]. Spark plasma sintering (SPS) represents one of the most efficient techniques for obtaining highly dense pellets in short periods of time (see **Figure 1c**) compared to other methods such as cold or hot sintering[53–62]. In this work, the SPS temperature was varied (650, 665, 680 and 700 °C) to optimize the compacity of the pellets, and to study the quantities and compositions of the grains and inter-grain phases in correlation with the overall ionic conductivity performance of the pellets[54,56,63] (see the Experimental Methods section and **Figure S1b** in the Supporting Information). The compacity values of the sintered pellets calculated using the **Eq. S1** are in the 85 - 99 % range (**Figure S2**), increasing from its lowest value at a sintering temperature of 650 °C up to its highest value for a sintering temperature of 680 °C, before decreasing again at higher temperatures. From the top view SEM images shown in **Figure S2** the average grain size for LAGP powder is ~0.65 µm, which increases up to ~1.00 µm for the 700 °C sintered pellet.

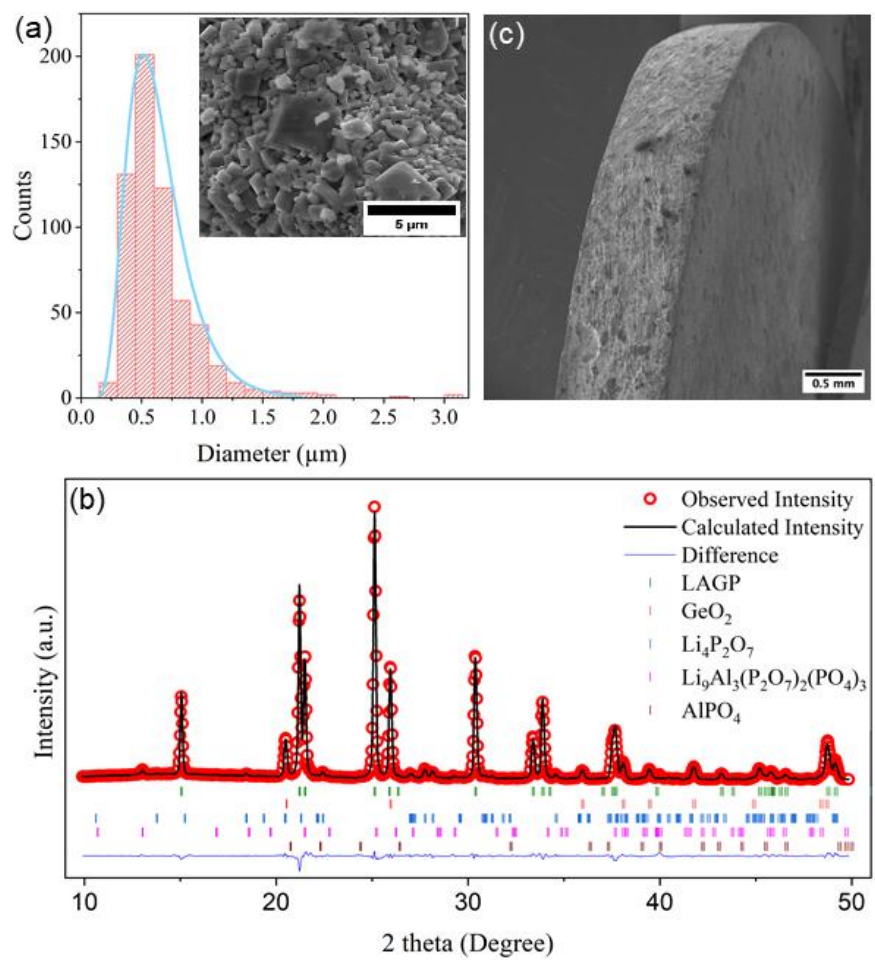

**Figure 1.** (a) Secondary electrons images and size distribution of the Li$_{1.5}$Al$_{0.5}$Ge$_{1.5}$(PO$_4$)$_3$ powder obtained by sol-gel method. The pink line is the histogram (N=612 particles) of the particle size distribution and in blue line is the normal fit of the size distribution (mean of 0.649 (±0.317) µm) (b) XRD Rietveld refinement analysis of the Li$_{1.5}$Al$_{0.5}$Ge$_{1.5}$(PO$_4$)$_3$ powder obtained by sol-gel method. (c) Cross-section secondary electron image of LAGP made pellet using Spark Plasma Sintering.

The ionic conductivities of the sintered pellets were measured using electrochemical impedance spectroscopy (EIS) over a temperature range spanning -30 - 60 °C. The measured Nyquist plots of the four pellets recorded at room temperature (RT) are shown in the **Figure 2a**. Each Nyquist plot is composed of two semicircles (at higher frequencies) followed by a linear tail at lower frequencies. The impedance data can be well fitted using a brick layer model (**Figure2a**) which consists of an instrumental resistance (R$_1$), two parallely connected resistance/CPE (constant phase element) couples connected in series representing the grain and grain boundaries, and a CPE representing the capacitive effect at the blocking electrodes (CPE4). The semi-circle (represented by R3 and CPE3) located at the higher frequencies was attributed to the Li$^+$ motion in the inter-grain region, whereas the second one (represented by R2 and CPE2) was attributed to the motion in the grains [26,29,49,50]. Grain and inter-grain phase ionic conductivity ($\sigma_g$ and $\sigma_{ig}$) were calculated based on the obtained resistances values using the **Eq. S2** and the overall ionic conductivity is the sum of grain and inter-grain ionic conductivity.

The RT impedance data of **Figure 2a** show that pellets sintered at 650 and 665 °C have much larger inter-grain resistivities (1,155.4 and 1,074.9 ohm.cm, respectively) in comparison to their 680 and 700 °C counterparts (195.7 and 322.6 ohm.cm, respectively), with the 680 °C pellet demonstrating the lowest inter-grain resistivity. Interesting enough, the bulk grain conductivity, contrary to inter-grain ionic conductivity, remains constant (to within experimental error) regardless the sintering temperature. As a result, the total ionic conductivity was monotonically increased with the sintering temperature until 680 °C, then was slightly decreased at 700 °C. A tendency towards higher conductivity was observed with higher compacity, except for the values obtained with 665 °C- and 700 °C-sintered pellets. This difference is due to a lower inter-grain resistivity present in 700 °C-sintered pellet (**Figure 2b**), which could originate from morphological and compositional changes in grain boundaries/inter-grain phases.

From the R2 and R3 resistance values summarized in **Table S1** obtained by fitting the impedance data recorded in the -30 - 60 °C range, the Arrhenius plots of **Figures 2c-e** associated with the grain (1/R2), inter-grain (1/R3) and total (1/R2+1/R3) ionic conductivities versus 1/T can be obtained. The slopes of the $\sigma$T vs 1/T Arrhenius plots allow a determination of the activation energies ($E_a$) of the ionic transport through the inter-grain and grain networks. Marginally smaller grain activation energies of ~0.42 eV are elucidated for the 680 and 700

°C samples with higher sintering temperatures, in comparison to grain activation energies of ~0.44 eV obtained from the samples prepared under the lower sintering temperature conditions (650 and 665 °C), thus suggesting lower energy barriers to ion migration are experienced for the higher sintered samples. In contrast, for the ion migration through the inter-grain region, the activation energies derived for the samples synthesized at the lower sintering temperatures exhibit marginally lower activation energies indicating a less disturbed ionic migration through the inter-grain region. This difference reflects the range of materials co-existing in the inter-grain regions and the evolving speciation that dominates these regions at the different sintering temperatures.

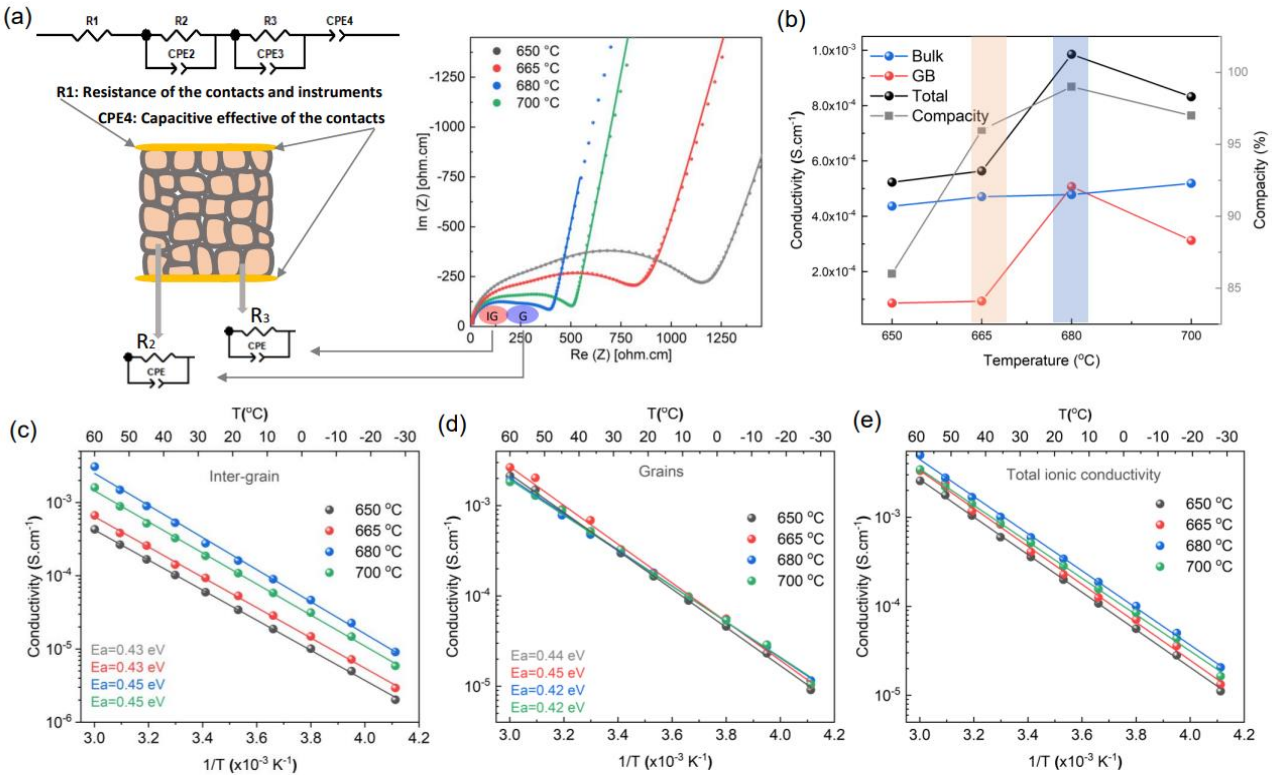

**Figure 2.** (a) Schematic representation of the equivalent circuit model and of the grain structure in the LAGP film and room temperature Nyquist plots of the LAGP pellets sintered at 650, 665, 680 and 700 °C; the fitted semi-circles of the grain and inter-grain regions are displayed. (b) Compacity and room temperature ionic conductivity of grain, inter-grain, and pellet values versus sintering temperatures. (c-e) Arrhenius plots of the (c) inter-grain, (d) grain and (e) total ionic conductivity values. The resistance and conductivity data are normalized by the specimen geometrical dimensions of the pellets.

To understand better the morphology and the composition of the inter-grain regions on the solid electrolyte conductivity, FIB-SEM tomography experiments ware carried out[64–66]. FIB-SEM tomography provides a 3D representation of the grain and inter-grain region at the micrometer scale, as well as allowing for a quantification of the phase fractions, tortuosity, and local connectivity. Moreover, as the level of contrast in the backscattered electron-SEM (BE-SEM) images can be associated with a given phase, the segmentation of the BE-SEM 3D volume allows to quantify and determine the spatial distribution of each phase present inside the SE. **Figure 3(a-d)** shows the BE-SEM images that were collected during the FIB-BE-SEM tomography acquisition of the pellets sintered at different temperatures. It is observed that in all the sintered pellets reveals the presence of an inter-grain region and pores surrounding the LAGP grains. The 3D-FIB-BE-SEM analysis shows that the inter-grain phase forms a connected network in all the LAGP grains which can strongly impact the ionic conduction. The presence of such an inter-grain phase located between the LAGP grains most likely constitute the main reason for a lower grain boundaries/inter-grain region ionic conductivity in the pellets sintered at lower temperature. The 3D volumes of the different phases (LAGP grains, inter-grain regions and voids) associated to the pellet sintered at 680 °C are shown in the **Figure 3(g-i)**. The 3D volumes of the pellets obtained at other temperatures and their associated phases (LAGP grains, inter-grain regions, voids, germanium oxide and Al-rich phase) are shown in the **Figure S3**. From 3D segmented phases, we can extract their relative fraction. Based on the 3D-FIB-BE-SEM analysis the low sintering temperature presents a higher fraction ratio of inter-grain region (spanning 13.8 − 21.4 % range) compared to high sintering temperature (spanning 13.8 − 21.4 % range). The lowest fraction ratio of the inter-grain region was observed for the 680 °C pellet, which also exhibited the lowest grain boundaries contribution suggesting a link between the fraction ratio of the inter-grain phase and total ionic conductivity.

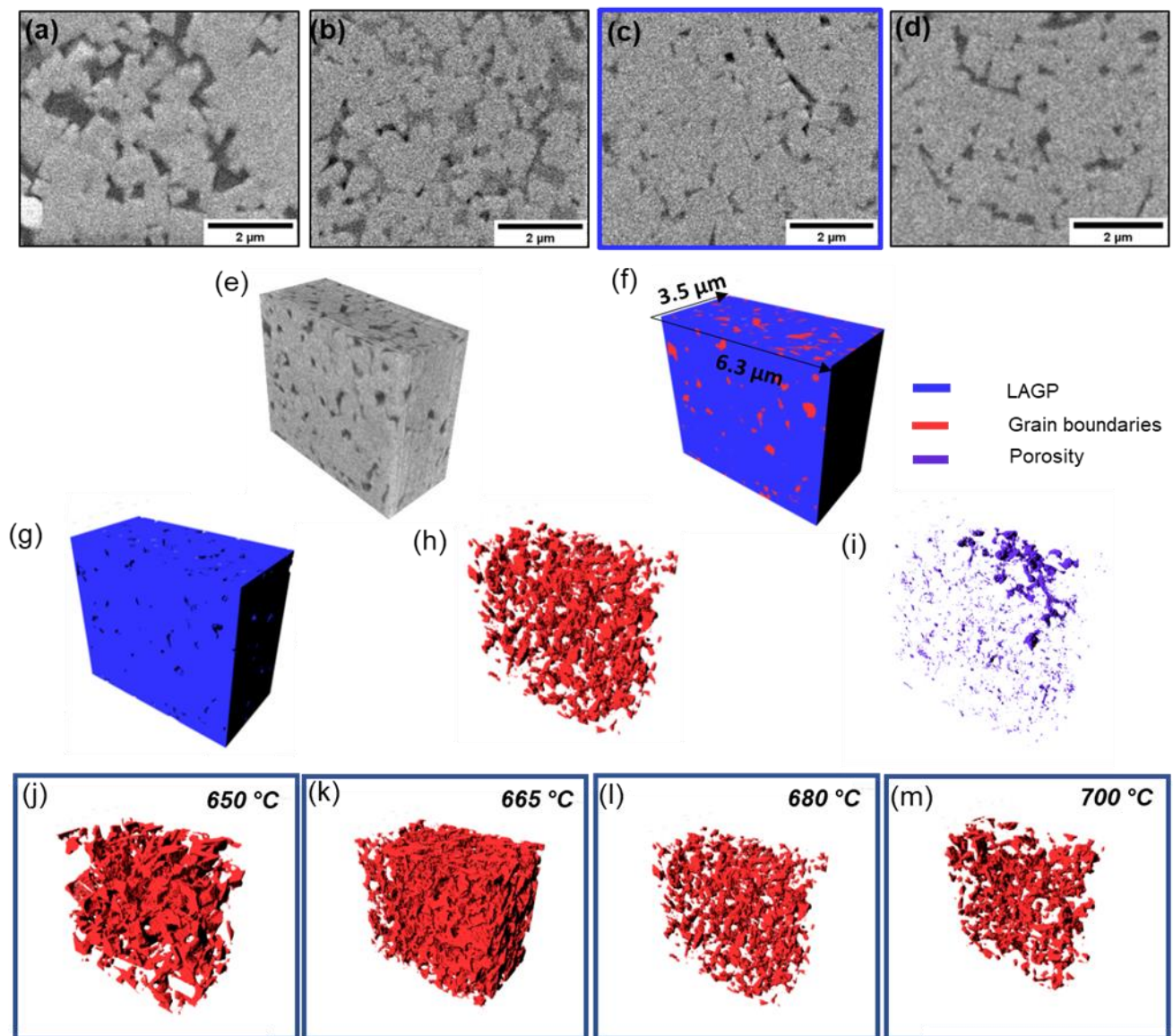

**Figure 3.** Backscattering electrons images from the FIB-SEM tomography of the pellets sintered at (a) 650 °C, (b) 665 °C, (c) 680 °C, (d) 700 °C. FIB-SEM tomography on the solid state electrolyte pellet prepared by SPS at 680 °C where it shows e) 3D volume reconstructed from a series of BE electron images, (f) 3D volume after segmentation, based on BE electron images of the three phases shown in g to I, (g) LAGP grains, h)Inter-grain phase, i)pores and voids. Reconstructed 3D volumes of inter-grain phase from LAGP pellets sintered at (j) 650°C, (k) 665 °C, (l) 680 °C and (m) 700 °C.

To understand better the ionic pathways in the SE, the tortuosity values of the LAGP and inter-grain phases are extracted from the 3D volumes of grains and inter-grain. The tortuosity represents how direct the 3D pathway is from a point to another in the volume. In terms of ionic conductivity, a tortuosity value of 1 represents a direct path, a larger tortuosity value a less direct paths (more tortuous) between the two sides of the pellets[67]. The tortuosity maps are given in **Figure S4** and their values in **Table 1** with the associated phase fraction.

The tortuosity values of the inter-grain regions were decreased from 6.2 to 2.2 when the sintering temperature increased from 650 °C to 665 °C and become infinite at higher sintering temperatures. The finite value indicates the presence of a continuous pathway in the inter-grain regions. The infinite values for higher sintering temperatures indicate the inter-grain region does not form a continuous path for the conduction of Li⁺. The tortuosity values for the grain remain close to 1 except for the 665 °C sintered pellet due to the presence of a dense inter-grain network that prevent a continuous connection of the LAGP grains. The similar tortuosity value of the grain is to put in relation with the similar activation energy measured for the grains by impedance measurements. The lower tortuosity values of the 650 °C and 665 °C pellets might explain the lower activation energies measured for the inter-grain regions for these samples by impedance, *i.e.*, a more continuous inter-grain pathways decreases the energy barrier for the Li⁺ to propagate through the inter-grain network. The tortuosity explains the difference between the activation energies but fails to explain the increase of the ionic conductivity in the pellet sintered at 680 °C.

**Table 1:** Phases fractions, tortuosity and compacity values for the four pellets sintered at 650 °C, 665 °C, 680 °C and 700 °C.

| Sintering temperatures | 650 °C | 665 °C | 680 °C | 700 °C |
|---|---|---|---|---|
| Grain phase fraction (%) | 78.12 | 78.53 | 94.07 | 92.81 |
| Inter-grain phase fraction (%) | 13.85 | 21.43 | 5.51 | 6.96 |
| Porosity fraction (%) | 0.02 | 0.74 | 0.42 | 0.00 |
| Others phase fraction (%) | 0.01 | 0 | 0 | 0.24 |
| Tortuosity: Grain | 1.09 | 1.44 | 1.03 | 1.04 |
| Tortuosity: Inter-grain | 6.15 | 2.22 | inf | inf |
| Compacity (%) | 86 | 96 | 99 | 97 |

To interrogate the inter-grain region on a local scale and its influence on the ionic conductivity, BE-SEM images and EDX-SEM analysis were performed (**Figure S5 and Figure S6**) at the end of the tomography process allowing a cross-section view of the pellet. BE-SEM images were formed from a signal proportional to the locally averaged atomic number (Z) of the sample, using such a Z sensitive imaging mode, and in combination with an EDX-SEM dataset acquired at the same locations, each level of contrast in the BE-SEM image can be attributed to a given composition. Using the BE-SEM images, a segmentation process was carried out (**Figure S5a**) to separate the different phases presented in the analyzed area of each sintered pellets. Thus, the five different levels of contrast in BE-SEM images were attributed to the $GeO_2$, LAGP, $AlPO_4$, inter-grain phase and voids with decreasing level of intensity from white to dark for the 650 °C sample. Further, using a python segmentation code (See details in Supporting Information) the average atomic percent of each phase present in the sample was obtained. The relative at% of each element (except Li) in the grain and inter-grain regions was listed in **Table S2.** Germanium concentration is found in a significantly lower value for the inter-grain phase in all the analyzed pellets compared to the grains in which its values are closer to the theoretical LAGP. Normalized average EDX spectra of segmented phases (**Figure S5f**) confirms that germanium is found in a lower concentration for the inter-grain phase (blue). Interestingly, phosphorus concentrations are lower than the theoretical values for the grains in all the analyzed pellets, but its values for the inter-grain are close to the theoretical LAGP. Aluminum concentration exhibited high variations according to the sintering temperature in the grains and inter-grain region. At 650 °C similar aluminum concentration was found in grains and at the inter-grain region. At 665 °C, a higher aluminum concentration was observed for the inter-grain region and at higher sintering temperatures (680 and 700 °C) aluminum was detected in a higher percent for the LAGP grains. These changes may affect the local $Li^+$ composition and alter the Li$^+$ conductivity, which could explain the significant differences in the inter-grain resistivities in pellets sintered at different temperatures. Oxygen was found in a higher concentration than LAGP theoretical values in both grains and inter-grain regions

The different crystalline phases present in the pellets are identified on a millimeter scale pellet using XRD (**Figures S7**) and the relative compositions of each phase were calculated by Rietveld refinement and were given in **Table S3**. The wt % of the main phase, LAGP, varied from 73 % to 82 %, depending on the sintering temperature. The second most dominant phase is $GeO_2$ about ~10 %, which had micrometer-scale grains according to BE-SEM imaging (**Figure S8**). Aside from its impact on the tortuosity pathway, $GeO_2$ phase is neither expected to contribute or be detrimental to the ionic conductivity as it produces non-conducting islands that do not impact the overall conduction pathways[68]. The third most dominant phase was $Li_4P_2O_7$ varying from ~10 % for pellet sintered at 650 °C to less than ~4 % for the 680 °C sintered pellet. [70]The presence of $Li_4P_2O_7$ in a mixture with other solid electrolyte such as in $Li_3PO_4$ was reported to induce an increase in the ionic conductivity by two order higher[39]. Interestingly, $Li_4P_2O_7$ or $Li_9Al_3(P_2O_7)_2(PO_4)_3$ were not observed by BE-SEM or EDX-SEM despite their 10.2 and 4.8 % at. concentration obtained from XRD measurements.

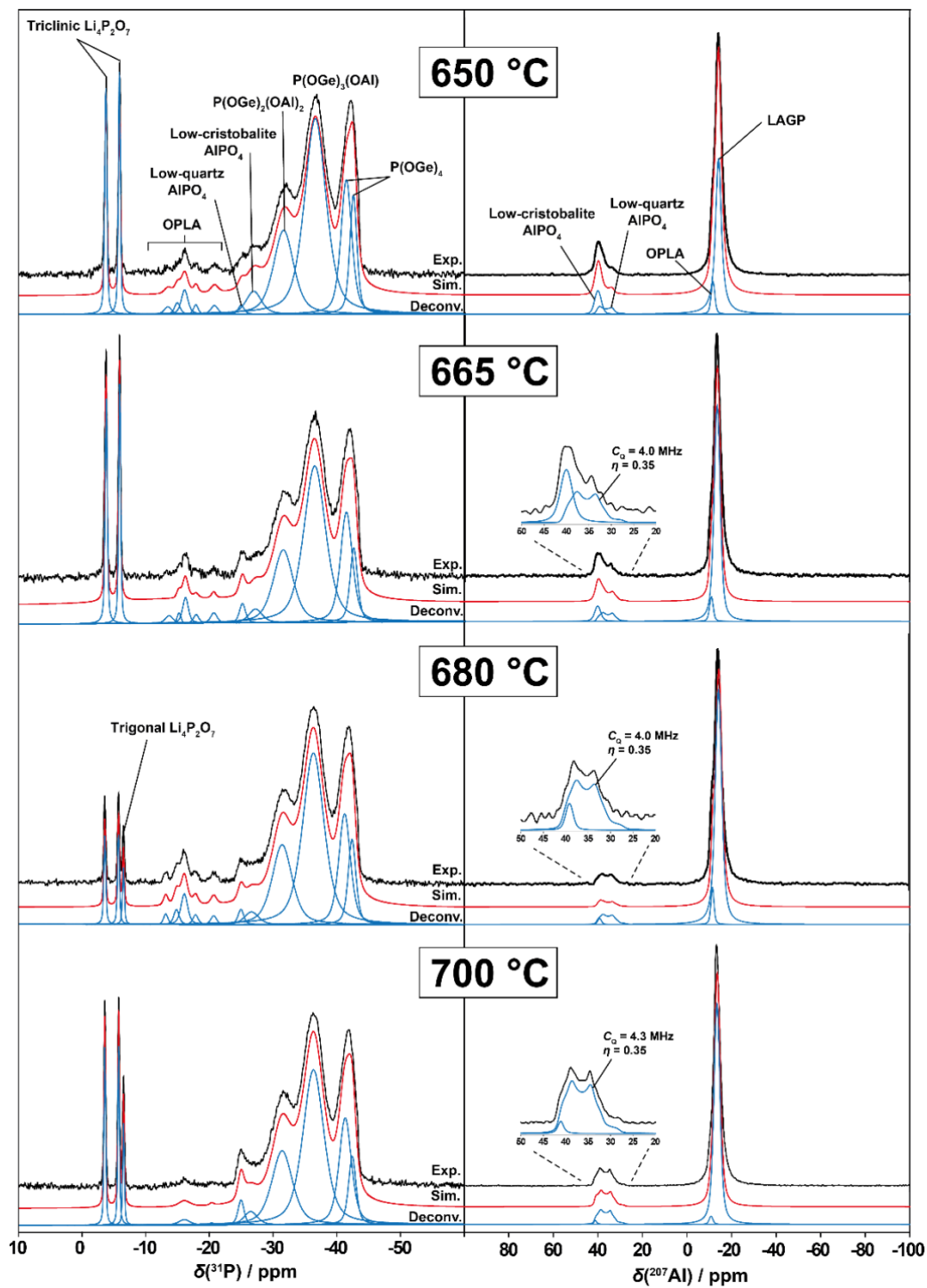

**Fig. 4** The solid state $^{31}P$ MAS NMR data (left, $v_0$ = 130.3 MHz, $v_r$ = 30 kHz) and $^{27}Al$ MAS NMR data (right, $v_0$ = 202.4 MHz, $v_r$ = 20 kHz) measured using single pulse methods from the four LAGP systems SPS sintered at temperatures of 650, 665, 680 and 700 °C. The long relaxation delays employed for the $^{31}P$ MAS NMR data ensures that P speciation is quantitatively estimated.

The solid state $^{31}$P and $^{27}$Al MAS NMR data displayed in **Figure 4** corroborate the variety of phases present within the SPS prepared materials, including the predominant LAGP phase (Li$_{1.5}$Al$_{0.5}$Ge$_{1.5}$(PO$_4$)$_3$), OPLA (Li$_9$Al$_3$(P$_2$O$_7$)$_2$(PO$_4$)$_3$), the triclinic and trigonal polymorphs of Li$_4$P$_2$O$_7$, and the low quartz and low cristobalite polymorphs of AlPO$_4$. The deconvoluted $^{31}$P NMR parameters for each preparation are summarised in **Table S4**; the relative intensities provide accurate quantitative estimates of the P speciation as these $^{31}$P MAS NMR data were acquired using 5 × the longest $^{31}$P $T_1$ values (*i.e.* those of the Li$_4$P$_2$O$_7$ polymorphs). The measured NMR parameters for the Li$_4$P$_2$O$_7$ and AlPO$_4$ polymorphs are in good agreement with previous studies by Raguž *et al.*[69] and Müller *et al.*[70], respectively. Furthermore, the observed chemical shifts for the LAGP and OPLA phases corroborate previously reported data[71–73], although a detailed structural characterization and assignment of the numerous $^{31}$P OPLA resonances has not yet been reported. All $^{31}$P and $^{27}$Al chemical shifts reported in **Table S4** have been further corroborated using GIPAW DTF calculations of

An analysis of the quantitative $^{31}$P MAS NMR data shown in **Figure 4** and **Table S4** is presented in **Figure S9.** This analysis reveals that only minor differences in composition exist between the 650 and 665 °C preparations for grain and inter-grain region (< 2% variation for each phase quantity). Starting with the 680 °C sample, a significant increase in the LAGP phase is observed, coinciding with a measured increase in the bulk grain phase and decrease in the grain boundary phase. Furthermore, the 680 °C preparation exhibits a marked increase in the OPLA inter-grain phase which is accompanied by an equally significant decrease in the Li$_4$P$_2$O$_7$ inter-grain phase. As reported above in **Figures 2a and b**, these changes to inter-grain speciation for the 680 °C preparation coincide with the lowest inter-grain resistance and highest conductivity.

The crystal structure of highly ordered OPLA (Li$_9$Al$_3$(P$_2$O$_7$)$_2$(PO$_4$)$_3$) possesses two chemically distinct P species with very similar GIPAW DFT predicted $^{31}$P chemical shifts occurring at $\delta$ −16.1 and −16.2 ppm (see **Table S4**). A close inspection of the OPLA region of the $^{31}$P MAS NMR spectrum shows that there are at least five resonances spanning a $\delta$ ~13 - ~21 ppm range representing the OPLA system, thus suggesting that the OPLA formed in the LAGP inter-grain regions under SPS conditions is highly disordered. Attempts to model the disorder using DFT calculations demonstrate that specific arrangements are energetically favourable and predict a distribution of $^{31}$P chemical shifts that agrees with the experimental $^{31}$P MAS NMR data. Disorder involving the loss of Li$_2$O (via the removal of adjacent Li$^+$ and O$^{2-}$ atoms in the crystal structure) thus forming Li and O vacancies:

(1) 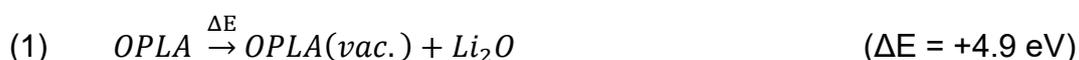  (ΔE = +4.9 eV)

yields a realistic total energy of formation and a resultant calculated $^{31}$P chemical shift distribution ranging from $\delta$ −12.6 - −22.8 ppm which is in excellent agreement with the experimental $^{31}$P MAS NMR data. Other descriptions of disorder creating Li deficiency involved the substitution of Ge$^{4+}$ for Al$^{3+}$ accompanied by the removal of an adjacent Li$^+$ atom to maintain charge balance:

(2) 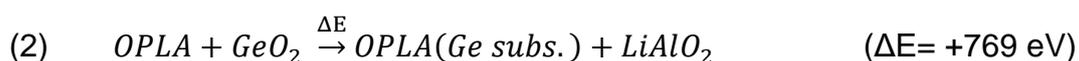  (ΔE= +769 eV)

These scenarios proved to be energetically unstable resulting in a more distributed $^{31}$P chemical shift range of $\delta$ −12.1 - −35.3 ppm which was not in agreement with the experimental data. The increased concentration of disordered/Li-deficient OPLA may be aiding ion migration through the inter-grain regions, consequently reducing the inter-grain resistance and the increasing the inter-grain and overall conductivities.

Accompanying the dominance of the OPLA phase in the inter-grain regions of the 680 °C sample is the significant reduction of the Li$_4$P$_2$O$_7$ phase. The amount of the inter-grain Li$_4$P$_2$O$_7$ phase is minimized in this preparation, and a partial phase transformation from the triclinic to the trigonal polymorph is also observed in a ratio of triclinic: trigonal ≈ 2 : 1. The measured

$^{31}$P $T_1$s of these $Li_4P_2O_7$ polymorphs are extremely long (>650 s), and the very narrow linewidth-at-half-height indicate long $^{31}$P $T_2$s and high short range and long range structural order. These characteristics suggest low Li ion migration through the $Li_4P_2O_7$ polymorphs within the room temperature regime, thus delivering an inefficient relaxation mechanism(s) to the P speciation within the frameworks shown in **Figure S10**. An increased OPLA content within the inter-grain regions of the 680 °C sample would offset the inefficient Li ion migration through the co-existing $Li_4P_2O_7$ phases and induce higher conductivity.

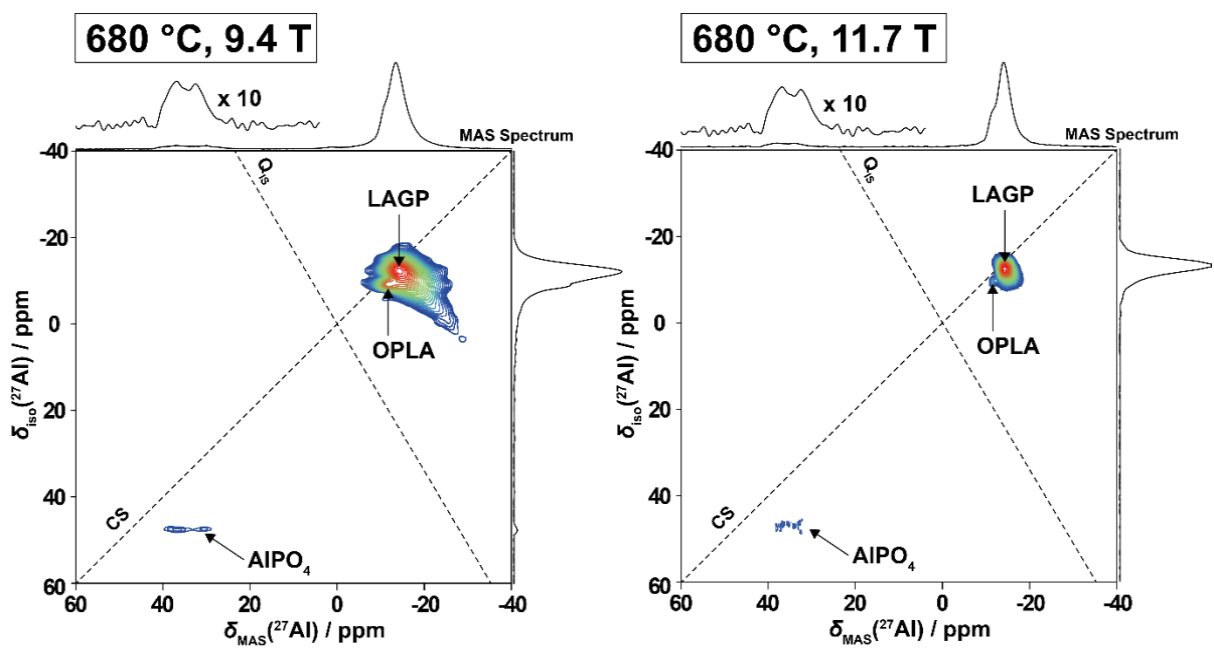

**Fig. 5** $^{27}$Al MQMAS NMR data for the 680 °C sample acquired at magnetic field strengths of 9.4 T (left) and 11.7 T (right).

The 1D $^{27}$Al MAS NMR data shown in **Figure 4** (all samples) and the 2D $^{27}$Al MQMAS NMR data presented in **Figure 5** (680 °C sample only) indicate the presence of Al speciation within the LAGP, OPLA and $AlPO_4$ phases. As expected from the substitutional and positional disorder characterising the LAGP ($Li_{1.5}Al_{0.5}Ge_{1.5}(PO_4)_3$) structure emanating from distributions of $P(OGe)_4$, $P(OGe)_3(OAl)$ and $P(OGe)_2(OAl)_2$ framework moieties. The contours 2D $^{27}$Al MQMAS NMR data associated with LAGP phase exhibit broad distributions along both the chemical shift (*CS*) and quadrupolar induced shift (*Q$_{IS}$*) axes, indicating large dispersions in both isotropic chemical shifts ($\delta_{iso}$) and quadrupolar coupling constants ($C_Q$) resulting from this disorder. In contrast, the $AlPO_4$ (low quartz polymorph) shows a horizonal contour indicative an Al species in a highly ordered environment where a single second order quadrupole broadening influences the $^{27}Al$ resonance (*i.e.* no distributions). In situations where structural disorder induces featureless resonances from quadrupolar nuclei (*i.e.* where no direct simulation of the resonance is possible to measure NMR parameters such as $\delta_{iso}$, $C_Q$ and $\eta_Q$), graphical methods involving variable field MAS NMR measurements can be employed to elucidate NMR parameters such as $\delta_{iso}$, and the quadrupole product $P_Q = C_Q\sqrt{(1 + \eta_Q^2/3)}$ which is a close approximation of the quadrupole coupling constant $C_Q$. The results of this analysis are presented in **Figure S11** demonstrating that very little change in the quadrupole parameters for the LAGP, OPLA and $AlPO_4$ phases is induced by the SPS sintering temperature. The quadrupole parameters are very sensitive indicators of structure and structural change; hence, the results of **Figure S11** suggest minimal structural change within each specific phase is experienced throughout this suite of preparations. The only prominent feature that changes throughout, and that which separates the 680 °C sample from the other preparations, is the amount of OPLA generated in the inter-grain regions as reported by the quantitative $^{31}P$ MAS NMR (see **Figure 4** and **Table S4**) and SEM-EDX analyses (See Figure S6). This pivotal feature is also reflected in the $^{27}Al$ MAS NMR data of **Figure S12**; however, reliable quantitative estimates are more difficult to derive from these data.

**Conclusions and Perspectives**

Spark plasma sintering (SPS) has been used to obtain highly compact and densified LAGP pellets. A range of sintering temperatures have been implemented to investigate the role of morphologies and compositions of the grains and inter-grain region on the ionic conductivity. The predominant finding from this work showed that the bulk ionic conductivity is constant throughout the different sintering temperatures, and that the main difference in the total ionic conductivity was induced by the ionic conductivity at the inter-grain region. The pellets sintered at a lower sintering temperature (650 and 665 °C) exhibited a higher resistance at the inter-grain regions compared to material produced at higher sintering temperatures (680 and 700 °C). The total ionic conductivity was observed to increase with the sintering temperature until 680 °C, followed by a small decrease at 700 °C. Even though sintering at 665 °C and 700 °C created similar compacity values, the latter had higher overall ionic conductivity indicating that morphological and compositional changes in the inter-grain region influence the ionic conductivity. 3D-FIB-SEM tomography revealed the presence of the inter-grain region, creating a network around the LAGP grains. The inter-grain region network was found in a lower proportion for the samples sintered at higher sintering temperature compared to lower sintering temperature, with 680 °C sample exhibiting the lowest inter-grain region and the highest ionic conductivity. Furthermore, along the morphology of the inter-grain region, their composition can explain the observed differences in the ionic conductivity. Cross-section segmented SEM-EDX analyses showed that the inter-grain phase contains significantly lower concentrations of Ge and a higher concentration of P compared to the bulk LAGP grain material. XRD and $^{31}$P MAS NMR analyses indicated the presence of $Li_4P_2O_7$ and OPLA ($Li_9Al_3(P_2O_7)_2(PO_4)_3$) phases in the overall LAGP pellet composition in concentrations up to ~10.2% and ~4.8 %, respectively; however, both phases were not explicitly identified by the BE-SEM or EDX-SEM analyses on the sintered pellets, indicating their presence in the inter-grain region. Based on the 3D-FIB-SEM tomography analysis, EDX-SEM, XRD and [31]P MAS NMR analysis, we concluded that the inter-grain phase presented like a network around the LAGP grains is a mixture of $Li_4P_2O_7$ and OPLA phases. Quantitative [31]P MAS NMR measurements demonstrate that optimal ionic conductivity for the LAGP system is achieved for the 680 °C SPS preparation when the disordered OPLA phase dominates the inter-grain region composition with reduced contributions from the highly ordered $Li_4P_2O_7$ phases, while the [27]Al MAS NMR data reveals that minimal structural change is experienced by each phase throughout this suite of sintering temperatures.

## Experimental Methods

### *Synthesis of the solid electrolyte*

The synthesis of the solid electrolyte, $Li_{1.5}Al_{0.5}Ge_{1.5}(PO_4)_3$, was done via a sol-gel method similar to the one described here[68]. The stoichiometric amount of $Al(OC_4H_9)_3$ (Sigma Aldrich) was dissolved in 100 mL THF solution for 2h. Using a syringe $Ge(OC_2H_5)_4$ (Sigma Aldrich) precursor was added drop by drop in the THF solution and then mixed together at 50 °C. In 100 mL water solution at 60 °C $CH_3COOLi$ and $NH_4H_2PO_4$ precursors were dissolved for 2 hours. After the dissolution of the precursors using a pipette the water solution is added drop by drop in the THF solution and let it overnight at 70 °C for the evaporation of the solvent. The powder obtained is grounded in an agate mortar and then placed for a heat treatment using a heat rate of 1 °C/min. at 850 °C for 2 hours for obtaining the LAGP phase.

### *Spark Plasma Sintering*

An FCT GmbH HPD10 Spark Plasma Sintering with the chamber in a glovebox was used to obtain compact LAGP pellets. The solid electrolyte powder was packed in a 10 mm (inner diameter) graphite die which was previously inside covered it with graphite paper. A pressure of 8 kN was applied on the pellets to compress the powder and a heating rate of 1 °C/s until the maximum temperature of heating, followed by a plateau of 5 minutes at the maximum temperature and then a controlled cooling within 5 minutes at the room temperature (total sintering time :27 min). The obtained pellets were polished to remove the graphite paper.

### *XRD analysis*

Each pellet obtained by SPS at different temperatures was grinded in a mortar and pestle to obtain a homogeneous powder. The obtained powder was characterized by X-ray diffraction using a D8 Bruker with Cu Kα radiation (λ1 = 1.5406 Å, λ2 = 1.5444 Å). The patterns were collected between 2θ = 10° -50° for 12 hours with a step of 0.02. Fullprof Suite was used to perform profile matching and Rietveld refinement in our samples. For the XRD analyses the following space groups and CIF file numbers of the different detected phases were addressed:

| Phase | Space group | CIF file number |
|---|---|---|
| $Li_{1.5}Al_{0.5}Ge_{1.5}(PO_4)_3$ | R3-c | 263763 |
| $GeO_2$ | $P3121$ | 66594 |
| $Li_4P_2O_7$ | $P$-1 | 59243 |
| $Li_9Al_3(P_2O_7)_2(PO_4)_3$ | $P$-3$c$1 | 50597 |
| $AlPO_4$ | $P3121$ | 9641 |

***Electrochemical impedance spectroscopy***

A gold layer of 100 nm was deposited on both sides of the pellets to insure a better contact with the blocking electrodes. The AC impedance spectra was recorded using an MTZ-35 (BioLogic) frequency response analyzer from a frequency range of 1 Hz to 10 MHz with an excitation signal of 5 mV coupled with an Intermediate Temperature System furnaces (ITS, BioLogic) temperature controller. The impedance data were treated using Zview software and Brick Layer Model (BLM) as equivalent circuit.

***Focused Ion Beam-Scanning Electron microscopy tomography***

ZEISS Gemini 2 Crossbeam 540 was used for the FIB-SEM tomography. An interest area of 6 µm x 5.5 µm was designated for all the solid electrolytes pellets. The interest area was chosen based on the top view images to avoid the presence of large secondary phases such $GeO_2$ and $AlPO_4$ and mostly focus on LAGP grains morphology. A platinum layer of 500 nm was deposited with the electron beam using a current of 2000 pA and another platinum layer using the focused ion beam of 600 nm with a current of 100 pA for the protection of the sample again the beam damage. To define the interest area, A double milling with 30 kV, 15 nA and 30 kV, 700 pA was realized to define the volume of interest to analyze. The automatic slice milling in the volume was realized using a current of 30 kV, 100 pA with a slice thickness of 21 nm and a depth of 8 μm. For the recording imaging conditions, we used 5 kV and 100 pA in order to avoid the damage in the solid electrolyte by a high current probe. Tilt compensator was set to 36 degrees to avoid the distortions which are induced in the image since the images are recorded when the sample is tilted to 54 degrees. Dynamic focus option was used to insure the focus from the top to the bottom of the recorded images.

### *Alignment, Segmentation, and 3D volumes reconstruction*

ImageJ software was used for the alignment of the images (Linear alignment with SIFT (Translation mode)) and the Trainable Weka Segmentation package was used for the segmentation of the images (Random Forest algorithm). The volumes were reconstructed using the software Dragonfly.

### *Solid State Nuclear Magnetic Resonance*

All $^{31}$P MAS NMR measurements were performed at 11.7 T using a Bruker Avance III spectrometer operating at a $^{31}$P Larmor frequency of 202.4 MHz. Rotational (MAS) frequencies of 30 kHz were achieved using a Bruker 2.5 mm HXY probe. Data were acquired at room temperature employing single pulse experiments which used π/8 pulses of 1 μs duration, and long recycle delay of 3600 s implemented to ensure the full relaxation of all $^{31}$P nuclei and $^{31}$P MAS NMR data. All $^{31}$P MAS NMR data were referenced against the IUPAC recommended reference of phosphoric acid ($H_3PO_4$) ($\delta_{iso}$ = 0.0 ppm) via a secondary solid reference of ammonium dihydrogen phosphate (ADP) ($\delta_{iso}$ = 0.99 ppm).

The $^{27}$Al MAS NMR measurements were acquired at using Bruker Avance III-500 (11.7 T), Bruker Avance III HD-400 (9.4 T) and Varian InfinityPlus-300 (7.05 T) spectrometers operating at a $^{27}$Al Larmor frequencies of 130.3, 103.9 and 78.2 MHz, respectively. All measurements were performed at room temperature and at MAS frequencies of 20 kHz, using Bruker 3.2 mm HX probes at 11.7 and 9.4 T, and a Varian HX probe at 7.05 T. Non-selective (solution) π/2 pulse times of 18 μs were calibrated using 1.1 M AlNO$_3$ from which a selective (solid) π/2 pulses of 6 μs were associated. All 1D $^{27}$Al MAS NMR data were acquired using single pulse experiments employing selective 1 μs π/12 pulses, and a recycle delay of 10 s. The corresponding 2D $^{27}$Al MQMAS data were acquired at 11.7 and 9.4 T using a 3Q Z-filtered experiment consisting of a 4 μs multiple quantum excitation pulse, a 1.4 μs conversion pulse, a soft π/2 20 μs selective pulse, a 20 μs inter-pulse Z-filter delay, and a 5 s recycle delay. A total of 80 slices were acquired to generate the indirect dimension, and 192 transients were acquired per slice. All $^{27}$Al MAS NMR data were referenced against the IUPAC recommended reference of 1.1 M aluminum nitrate (AlNO$_3$) solution ($\delta_{iso}$= 0 ppm). The determination of the NMR parameters from featureless quadrupole dominated $^{27}$Al resonances is achieved by graphical presentation of variable field $^{27}$Al MAS NMR data. The $^{27}$Al apparent center-of-gravity shifts $\delta_{cg}$ are the sum of the field independent isotropic chemical shift $\delta_{iso}$, and the field-dependent second-order quadrupolar shift $\delta_{Q,iso}^{(2)}$ such that:

$$\delta_{cg} = \delta_{iso} + \delta_{Q,iso}^{(2)}(I,m)$$

where

$$\delta_{Q,iso}^{(2)}(I,m) = \left(\frac{3C_Q^2}{40\nu_0^2 I^2 (2I-1)^2}\right)[I(I+1) - 9m(m-1) - 3](1 + \frac{\eta^2}{3})$$

By measuring the $\delta_{cg}$ variation over multiple magnetic fields (i.e. by changing $\nu_0$), a determination ot the $\delta_{iso}$ (from the y intercept) and $P_Q$ (from the slope) parameters is achieved by linear regression of the above equation for $\delta_{cg}$. The $P_Q$ parameter:

$$P_Q = C_Q \sqrt{(1 + \frac{\eta^2}{3})}$$

is closely related to the quadrupolar coupling constant $C_Q$ by a factor of ~10 - 15%.

***Density Functional Theory (DFT) Calculations***

CASTEP version 17.21[74], a plane-wave pseudopotential DFT package was used to perform geometry relaxation calculations, NMR calculations, and energy calculations for all systems. The PBE generalized gradient functional was used to model the exchange-correlation energy term[75]. CASTEP's built-in "on-the-fly" C17 pseudopotential library was used for these calculations. All unit cell parameters were fixed for geometry optimizations. Atomic positions were relaxed up to a tolerance of 0.05 eVÅ$^{-1}$ in atomic forces for all systems. NMR parameters were computed using the GIPAW method[76,77]. For both geometry relaxation calculations and NMR parameter calculations, an electronic cut off energy of 600 eV for the plane-wave basis set was used. Structures for triclinic $Li_4P_2O_7$[78], trigonal $Li_4P_2O_7$[69], monoclinic $Li_4P_2O_7$[78], $Li_9Al_3(P_2O_7)_2(PO_4)_3$ (OPLA)[79], $Li_2O$[80], $GeO_2$[81], and $LiAlO_2$[82] were taken from associated references.

**Supporting information** is provided free-of-charge on the editor website.

## Acknowledgements

S.C. and A.D. acknowledge the region Hauts-de-France for the financial support. S.C., L.L.N. and M.D. acknowledge Nanyang Technological University (NTU) for financial support. FIB-SEM tomography analyses were performed at the Facility for Analysis, Characterization and Simulation (FACTS) at Nanyang Technological University, Singapore. J.V.H. acknowledges financial support for the solid-state NMR instrumentation at Warwick used in this research which was funded by EPSRC (grants EP/M028186/1 and EP/K024418/1), the University of Warwick, and the Birmingham Science City AM1 and AM2 projects which were supported by Advantage West Midlands (AWM) and the European Regional Development Fund (ERDF).

## Author contributions

S.C. performed synthesis of the solid electrolyte, electrochemical impedance spectroscopy analysis, XRD analysis, carried out the FIB-SEM tomography, analysis, segmentation, and 3D volume representations, EDX analysis using HyperSpy, prepared FIB lamellas and performed STEM analysis, as well as global analysis and composed the manuscript. L.L.N. helped with the acquisition of the FIB-SEM tomography. D.G.B. performed the NMR acquisition and analysis. T.T.N performed the tortuosity analysis. O.U.K. assisted with EIS and XRD Rietveld refinement analysis. B.F helped with EIS data acquisition and analysis, A.J. provided support and helped with the acquisition of the SEM images, J.-N.C. assisted with the Rietveld refinement analysis. V.S assisted with designed and assisted with the SPS experiments. J.V.H devised and supervised the NMR and materials modelling analysis. A.D. and M.D. supervised the work and J.V.H, A.D. and M.D. assisted with writing of the manuscript.

## Conflict of interest

The authors declare no competing financial interest.

# Supporting information

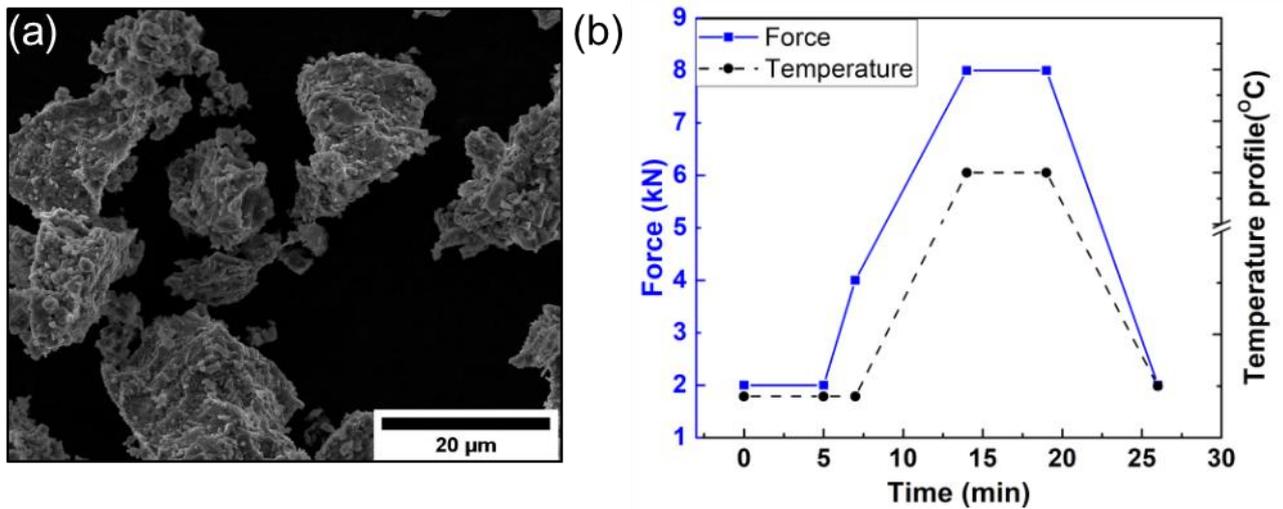

**Figure S1.** (a)Secondary electron image of LAGP particles (b) Spark plasma Sintering parameters used in order to obtain the compact pellets at 650 °C, 665 °C, 680 °C and 700 °C with a heating rate of 1 °C/s under a pressure of 100 MPa.

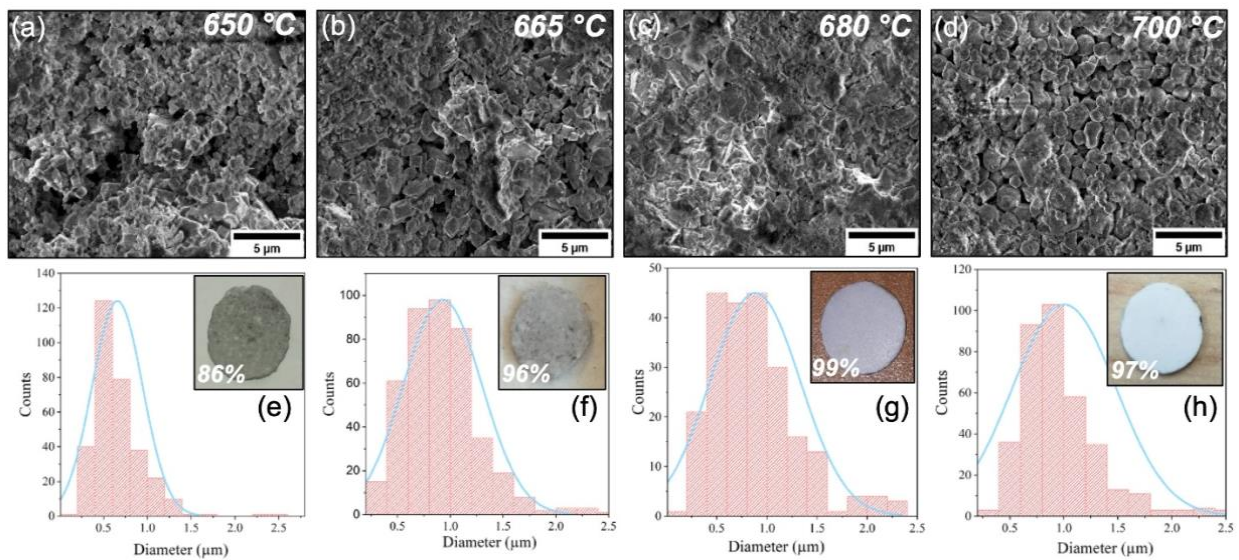

**Figure S2**. SEM Secondary electron images of LAGP surface pellets sintered at (a, e) 650 °C, (b, f) 665 °C, (c, g) 680 °C, (d, h) 700 °C and their associated particle size distribution histogram. The compacity values of each pellet is given in % in the insets.

Compacity values were calculated using the following equation:

$$\textbf{Compacity} = (m * \rho)/(\pi * r^2 * h) \qquad \textit{(Eq. S1)}$$

where:  h: thickness (cm)

m: mass of the pellet (g)

ρ: density (g/cm)

Grain ($\sigma_g$) and Inter-grain phase ($\sigma_{ig}$) ionic conductivities were calculated using the

following formula:

$$\boldsymbol{\sigma_g / \sigma_{gb}} = \frac{1}{Rg/R\,gb}\left(\frac{t}{A}\right) \qquad \textit{(Eq. S2)}$$

where: t= thickness of the pellet

A=surface area of the pellet

**Table S1**. Resistance values of the bulk and grain boundaries obtained after the fit with Brick-Layer model for the RT measurement and the values of the thickness/surface area rapport of the pellets.

|  | 650 °C | 665 °C | 680 °C | 700 °C |
|---|---|---|---|---|
| R bulk (ohm) | 218.1 | 156.7 | 227.3 | 237.7 |
| R grain boundaries (ohm) | 1060 | 762.4 | 206 | 435.9 |
| t/A (cm) | 1.09 | 1.41 | 0.95 | 0.74 |

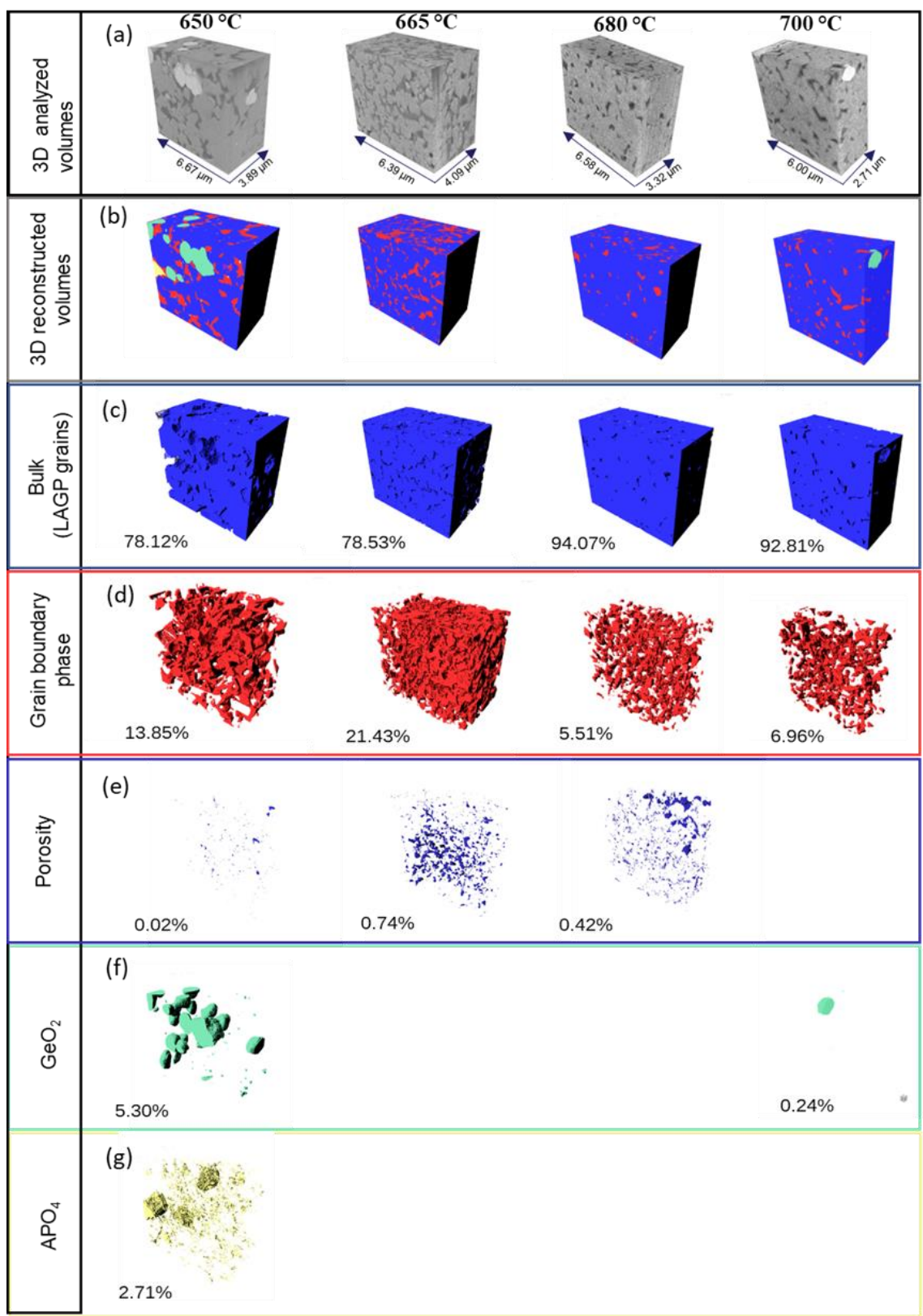

**Figure S3.** FIB-SEM tomography on the solid state electrolyte pellets prepared by SPS at 650 °C, 665°C, 680 °C and 700 °C where it shows (a) 3D volumes reconstructed from a series of BE electron images, (b) 3D reconstructed volumes after segmentation based on BE electron images of the five phases presented in the SE (c) LAGP grains, (d) pores and voids, (e) Inter-grain phase, (f) germanium oxide and (g) AlPO$_4$

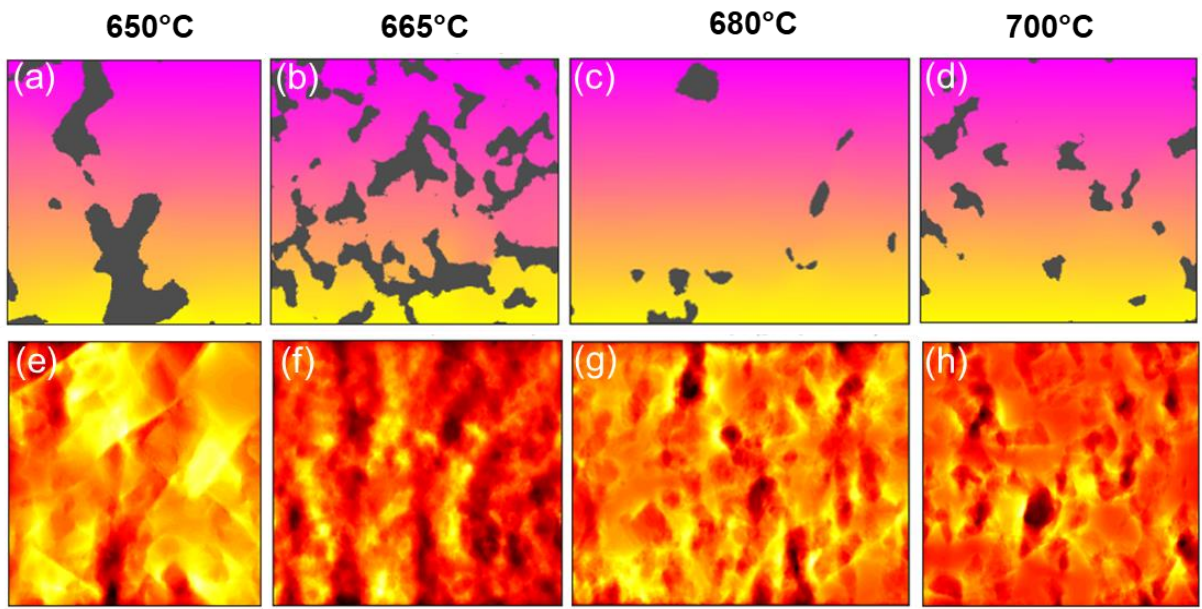

**Figure S4**. Tortuosity analysis on the four sintered pellets by Spark Plasma Sintering using the 3D volumes representation obtained by FIB-SEM tomography. **Table 1:** Atomic percent of Ge, Al, O and P detected by SEM-EDX for the grain and inter-grain phase

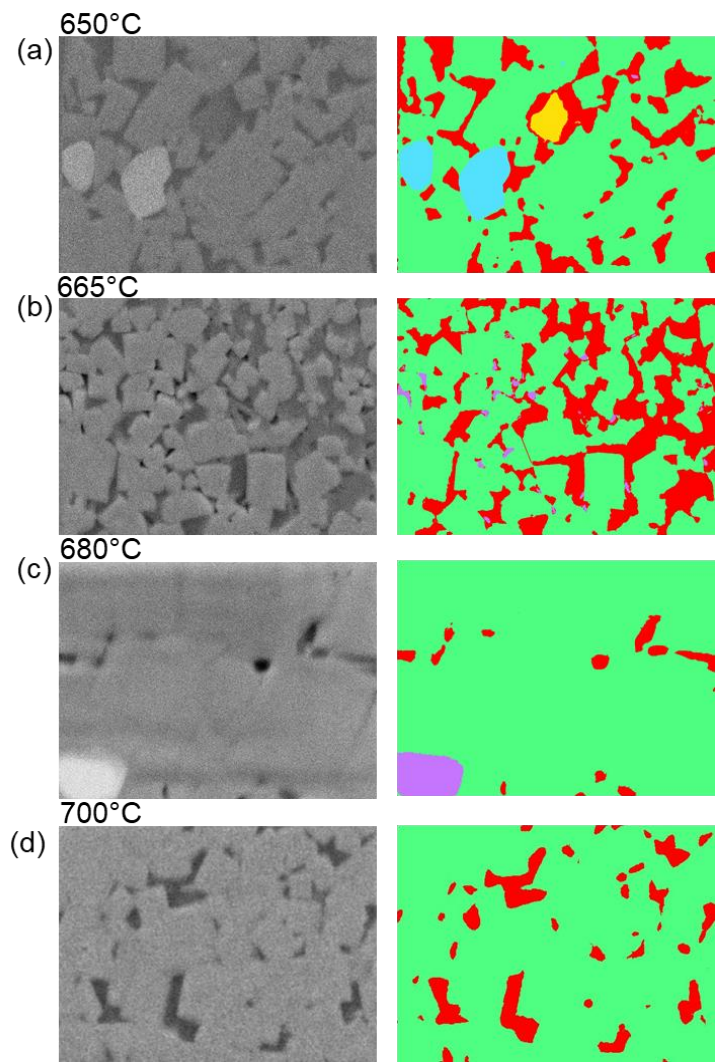

**Figure S5.** BE-SEM and segmented images used to obtain the concentration of each phase presented on our analyzed area.

Image segmentation process was done using the Fiji package, Trainable Weka Segmentation. For the 650 °C pellet on the segmentation image the colors are corresponding to the following phases. Green is attributed to LAGP grains, red to inter-grain phase, purple to pores and voids, blue to germanium oxide and yellow for the Al-rich phase. For the 665 °C pellet on the segmentation image green is attributed to LAGP grains, red to inter-grain phase and purple to pores. For the 680 °C pellet on the segmentation image green is attributed to LAGP grains, red to inter-grain phase and purple to germanium oxide. For the 700 °C pellet on the segmentation image green is attributed to LAGP grains and red to pores and voids.

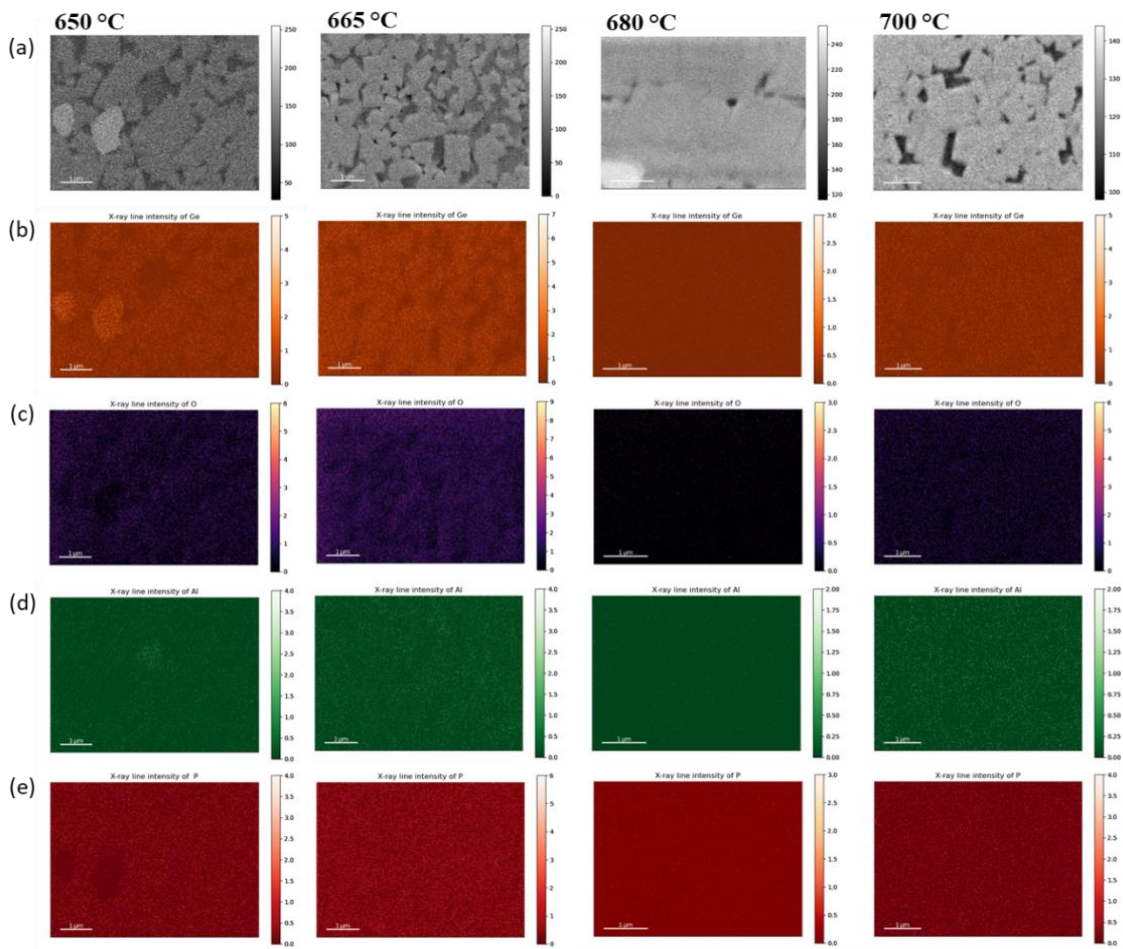

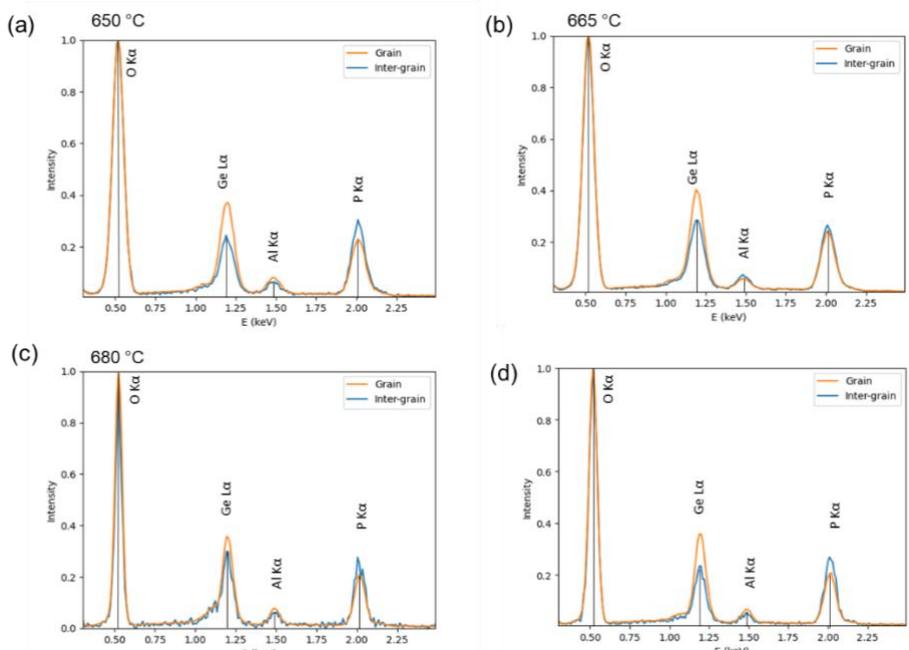

**Figure S6.** Cross-section SEM-EDX analysis performed on the solid state electrolyte pellets prepared by SPS at 650 °C, 665°C, 680 °C and 700 °C where it shows (a) the BE image corresponding the analyzed area and the X-ray line intensities of (b) germanium, (c) oxygen, (d) aluminum and (e) phosphorus. (f) Normalized average EDX spectra of grains and inter-grain phase obtained after a segmentation process where phases were separated based on their contrast.

**Table S2**. Atomic percent of Ge, Al, O and P detected by SEM-EDX for the grain and inter-grain phase

| Phase | at% Ge | at% Al | at% O | at% P |
|---|---|---|---|---|
| **Theoretical LAGP** | 8.81 | 2.92 | 70.60 | 17.65 |
| **650 °C grain** | 6.95 | 2.04 | 76.63 | 14.38 |
| **650 °C inter-grain** | 4.61 | 2.09 | 78.56 | 14.74 |
| **665 °C grain** | 8.03 | 1.56 | 78.25 | 12.16 |
| **665 °C inter-grain** | 5.83 | 2.54 | 77.97 | 13.66 |
| **680 °C grain** | 8.37 | 3.22 | 76.96 | 11.45 |
| **680 °C inter-grain** | 7.66 | 2.57 | 73.92 | 15.85 |
| **700 °C grain** | 7.92 | 2.43 | 78.29 | 11.36 |
| **700 °C inter-grain** | 5.04 | 1.81 | 78.07 | 15.07 |

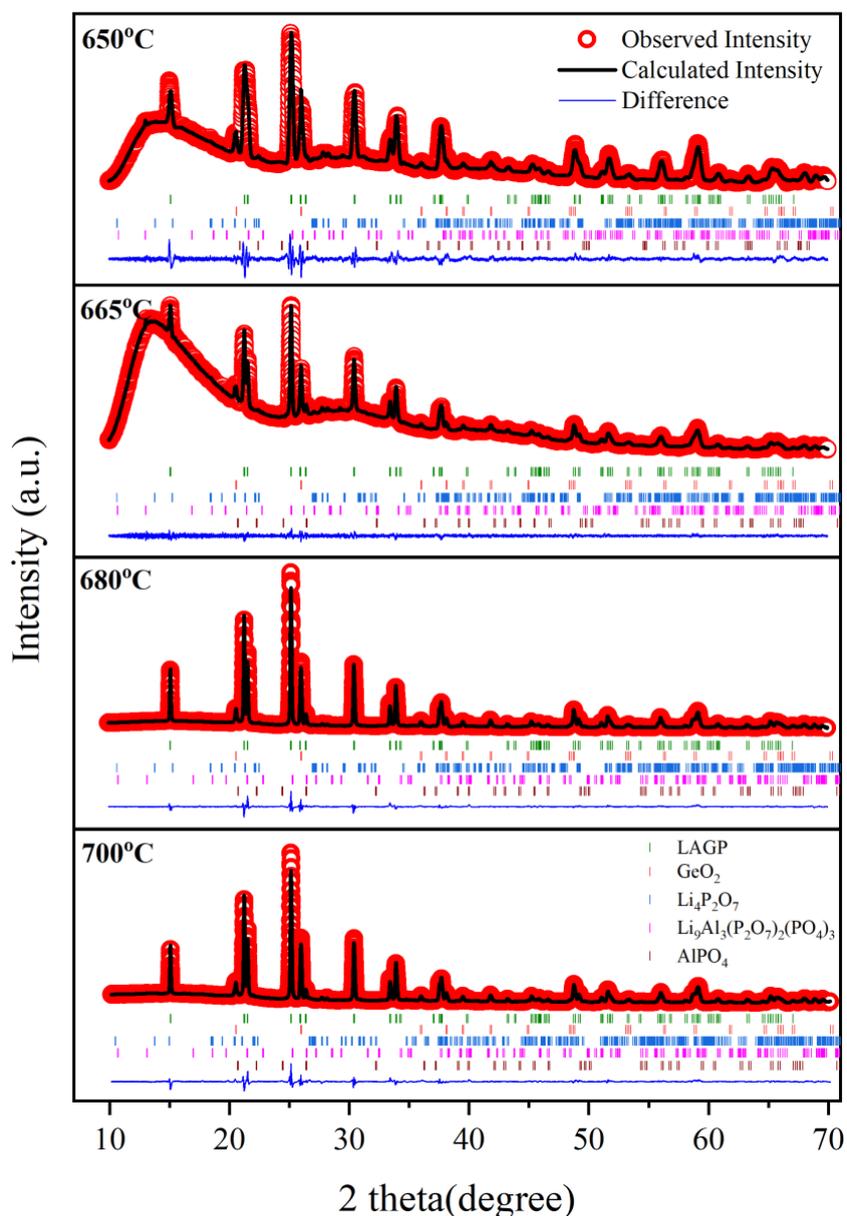

**Figure S7**. Rietveld refinement analysis of the four sintered pellets obtained by Spark Plasma Sintering.

**Table S3**. Rietveld analysis on the SPS sintered pellets at different temperatures.

|  | LAGP % wt. | GeO₂ % wt. | Li₄P₂O₇ % wt. | AlPO₄ % wt. | Li₉Al₃(P₂O₇)₂(PO₄)₃ % wt. | $R_P$ | $R_{wp}$ | $R_{exp}$ | S |
|---|---|---|---|---|---|---|---|---|---|
| **Powder** | 81.35(0.24) | 7.02(0.05) | 10.11(0.15) | 0.36(0.05) | 1.17(0.00) | 14.5 | 15.1 | 3.55 | 18.3 |
| **650 °C** | 73.33(0.78) | 13.45(0.15) | 10.20(0.40) | 0.50(0.07) | 2.52(0.22) | 22.1 | 17.3 | 7.54 | 5.28 |
| **665 °C** | 74.43(1.08) | 10.19(0.22) | 7.78(0.87) | 2.82(0.12) | 4.77(0.23) | 20.4 | 12.8 | 10.76 | 1.42 |
| **680 °C** | 82.27(0.21) | 10.30(0.05) | 3.83(0.14) | 3.61(0.06) | 0.00(0.00) | 10.2 | 10.4 | 2.15 | 23.5 |
| **700 °C** | 81.06(0.29) | 9.82(0.07) | 5.57(0.22) | 3.55(0.07) | 0.00(0.00) | 13.5 | 13.3 | 2.42 | 30.0 |

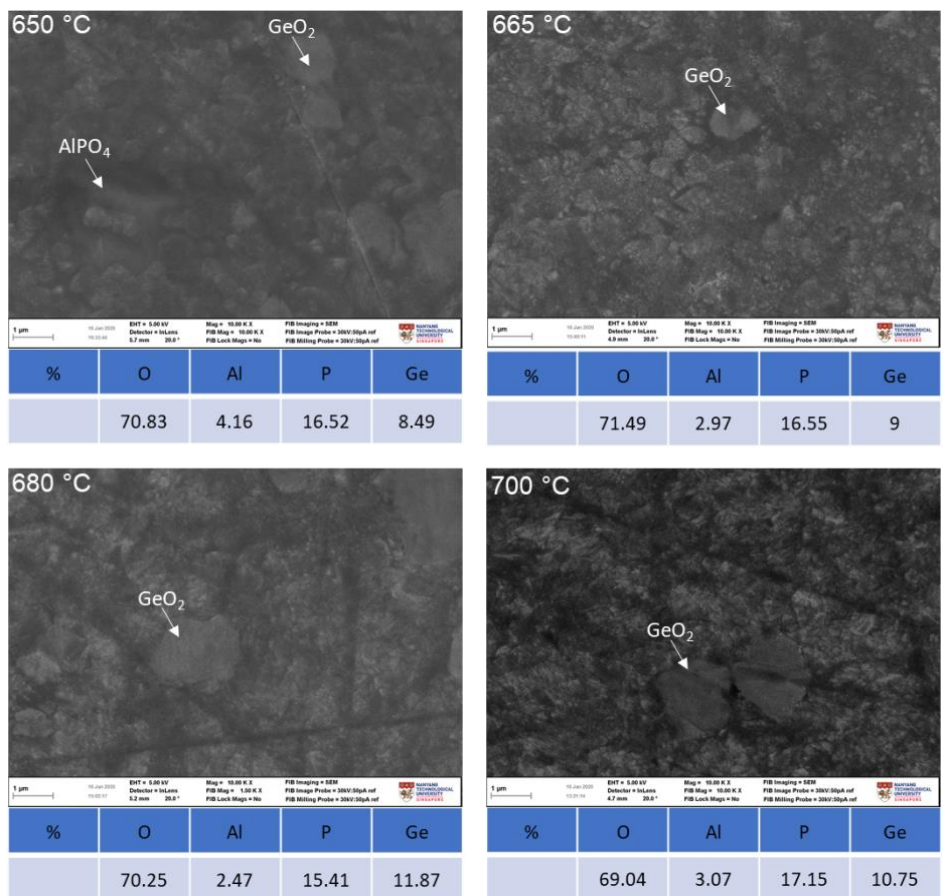

**Figure S8**. Selected top surface secondary electron images of LAGP pellets sintered at various temperatures where it shows the presence of $GeO_2$ and $AlPO_4$ inside of the solid electrolyte

**Table S4** Measured $^{31}$P and $^{27}$Al NMR parameters determined by lineshape simulation of the data displayed in Fig. 3. For comparison, the DFT-calculated $\delta_{iso}$ values for $Li_4P_2O_7$, $AlPO_4$ and OPLA phases are also shown.

| Phase | 650 °C | | 665 °C | | 680 °C | | 700 °C | | Calculated $\delta_{iso}$ (ppm) |
|---|---|---|---|---|---|---|---|---|---|
| | $\delta_{iso}$ (ppm) | Relative Intensity (%) | $\delta_{iso}$ (ppm) | Relative Intensity (%) | $\delta_{iso}$ (ppm) | Relative Intensity (%) | $\delta_{iso}$ (ppm) | Relative Intensity (%) | |
| $Li_4P_2O_7$ | -3.5 -5.7 | 5.1 5.0 | -3.6 -5.8 | 6.1 5.9 | -3.5 -5.7 -6.5 | 2.2 2.1 1.3 | -3.6 -5.7 -6.5 | 4.0 4.0 2.2 | -2.6 (Triclinic) -5.4 (Triclinic) -6.0 (Trigonal) |
| $Li_9Al_3(P_2O_7)_3(PO_4)_2$ | -13.2 -14.7 -16.0 -17.7 -20.7 | 0.5 0.7 1.6 0.5 0.6 | -13.5 -15.1 -16.1 -17.8 -20.6 | 0.7 0.4 1.8 0.6 0.8 | -13.2 -14.8 -16.1 -17.9 -20.7 | 0.6 1.0 0.6 2.5 0.6 | -16.1 -20.4 | 0.8 0.1 | -16.2 (Pristine) -12.6 – -22.8 (Li2O removed) -12.1 – -35.3 (Li removed, Ge subs. Al) |
| $AlPO_4$[a] | -24.9 -27.2 | 0.5 3.5 | -25.1 -27.1 | 1.4 2.2 | -25.0 -26.6 | 1.1 2.6 | -25.0 -26.5 | 2.0 2.3 | -24.5 (Berlinite) -23.3 (Low-cristobalite)[b] |
| Other NASICON Phases [P(OGe)4], [P(OGe)3(OAl)], [P(OGe)2(OAl)2] | -31.6 -36.4 -41.1 -42.3 | 16.8 41.9 14.7 8.5 | -31.5 -36.4 -41.4 -42.6 | 17.2 40.5 15.8 6.7 | -31.4 -36.4 -41.3 -42.4 | 18.8 43.5 15.6 8.0 | -31.4 -36.4 -41.4 -42.5 | 19.4 41.5 17.0 6.6 | |

[a] Possible overlap of low cristobalite $AlPO_4$ and LAGP phase.
[b] This shift was previously reported as $\delta_{iso} = -27.1$ ppm [1]

Overlap of the low cristobalite $AlPO_4$ and the LAGP resonances in the ~25 ppm region reduces the accuracy of the reported quantitative estimate of the low cristobalite $AlPO_4$ phase; however, the GIPAW DFT calculated $^{31}$P chemical shifts of these $AlPO_4$ phases confirms their appearance in this spectral region.

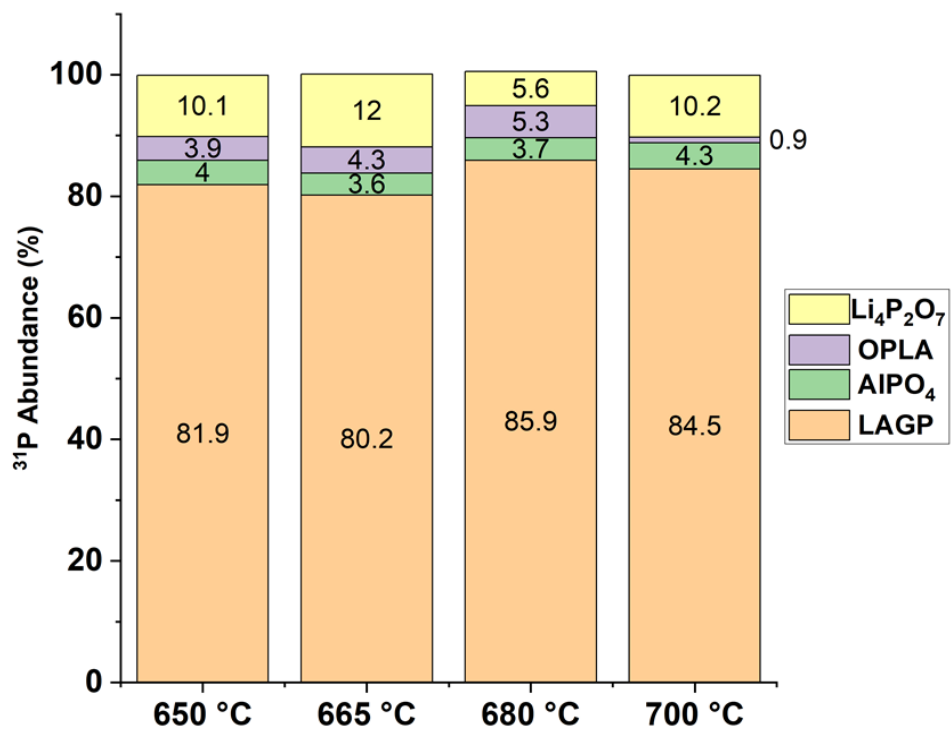

**Figure S9**. % of phases in the sintered pellets at various temperatures based on the relative intensities of $^{31}P$ abundance

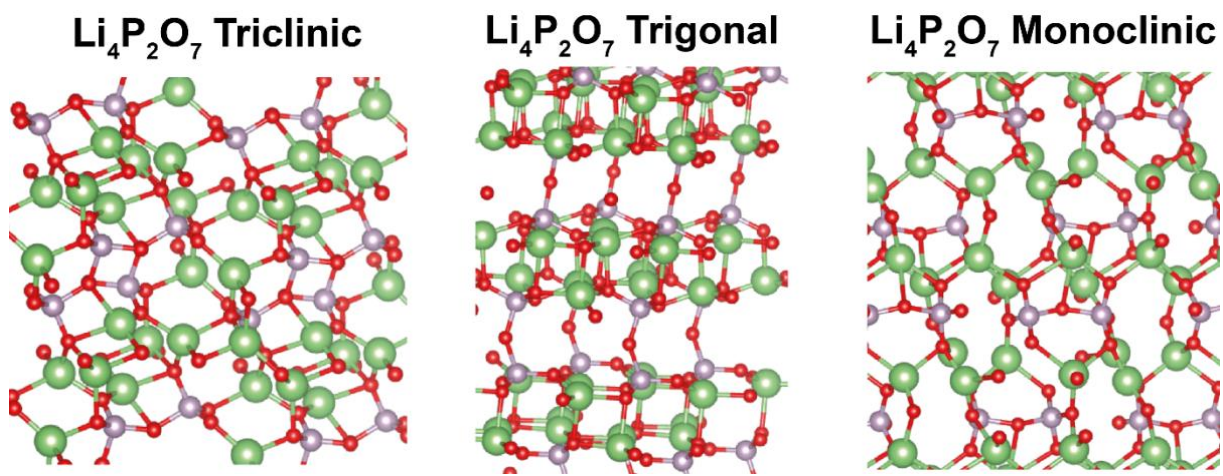

**Figure S10** Crystal structure diagrams of the triclinic, trigonal, and monoclinic polymorphs of $Li_4P_2O_7$.

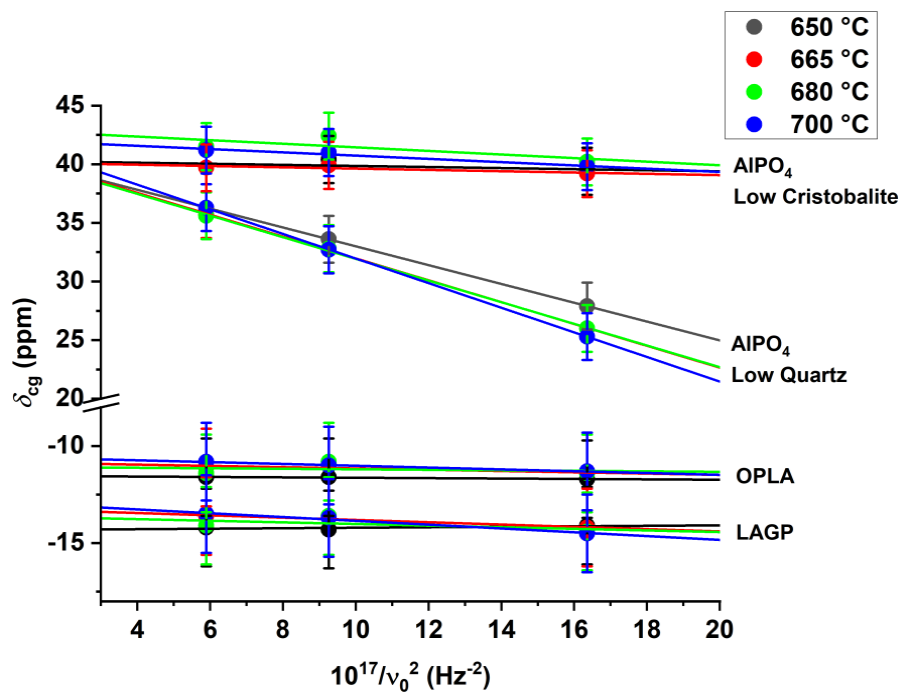

**Figure S11** Plot of the measured $^{27}Al$ centre-of-gravity shifts ($\delta_{cg}$) values for the four SPS synthesized LAGP materials as a function of the inverse squared Larmor frequency (magnetic field strength). Linear fitting of these data yields the isotropic chemical shifts ($\delta_{iso}$) determined from the y-intercept, and the quadrupole product values $P_Q = C_Q\sqrt{(1 + \eta_Q^2/3)}$ determined from the slope. More accurate NMR parameters for the low quartz $AlPO_4$ polymorph are presented in Figure 4 from direct simulation.

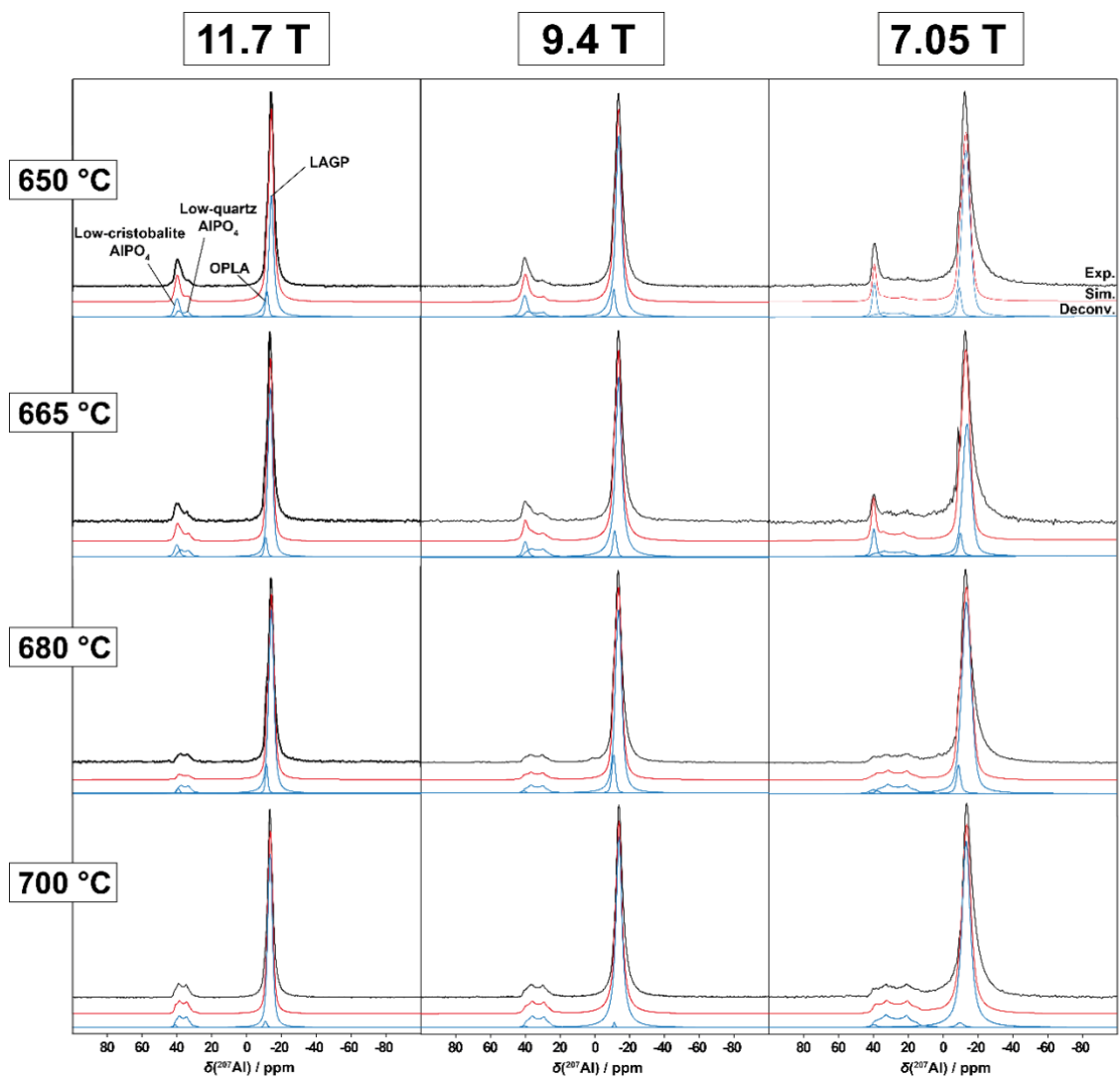

**Figure 12** Multiple field $^{27}$Al MAS NMR data of the four pellets, acquired at 11.7, 9.4 and 7.05 T. Lineshape simulations are shown in red, with individual component in blue. Determined NMR parameters are displayed in Table 3.

Github link for the Phyton code for the segmentation of EDX map.
https://github.com/sorinacretu/EDX_map-segmentation